# Power Spectrum Analysis for Optical Tweezers, II: Laser Wavelength Dependence of Parasitic Filtering, and how to Achieve High Band-Width

Kirstine Berg-Sørensen,[1, *] Erwin J. G. Peterman,[2] Tom Weber,[2, †] Christoph F. Schmidt,[2, ‡] and Henrik Flyvbjerg[3]

[1]*The Niels Bohr Institute, Blegdamsvej 17, DK-2100 Copenhagen Ø, Denmark*
[2]*Department of Physics and Astronomy, Faculty of Sciences,
Vrije Universiteit, De Boelelaan 1081, 1081 HV, Amsterdam, The Netherlands*
[3]*Biosystems Department and Danish Polymer Centre,
Risø National Laboratory, DK-4000 Roskilde, Denmark*
(Dated: March 7, 2006)

In a typical optical tweezers detection system, the position of a trapped object is determined from laser light impinging on a quadrant photodiode. When the laser is infrared and the photodiode is of silicon, they can act together as an unintended low-pass filter. This parasitic effect is due to the high transparency of silicon to near-infrared light. A simple model that accounts for this phenomenon (Berg-Sørensen et al., J. Appl. Phys., **93**, 3167–3176 (2003)) is here solved for frequencies up to 100 kHz, and for laser wavelengths between 750 and 1064 nm. The solution is applied to experimental data in the same range, and is demonstrated to give this detection system of optical tweezers a bandwidth, accuracy, and precision that is limited only by the data acquisition board's band-width and bandpass ripples, here 96.7 kHz, resp. 0.005 dB.

PACS numbers: 42.62.Be,42.70.-a,42.79.Pw,85.60.Dw,87.80.Cc

## I. INTRODUCTION

Photodiode-based detection systems are used in a number of modern techniques, ranging from detection of the cantilever deflection in atomic force microscopes [1, 2], over detection schemes coupled to optical tweezers [3–5, and references therein], to equipment used in high-energy physics particle detectors [6, 7]. When common Si-PIN diodes are used to detect 1064 nm laser light, over 50% loss of signal is seen for frequencies above approximately 10 kHz [8–11], and the loss is detectable from approximately 1 kHz [11].

This loss in signal power has a characteristic function similar to that of a low-pass filter, and is caused by light absorption in the n-layer of the diode. Its physics was explained and modeled mathematically in [10, 11]. The model was demonstrated to account fully for the phenomenon up to the 25 kHz Nyquist frequency used, and the resulting theory for the power spectrum agrees with the experimental spectrum to within the 1% stochastic error on the latter. The various degrees of loss in signal power was demonstrated in [12] for photo detectors made from different materials, including a specialized Si-diode, and for a range of laser wavelengths.

In the present paper, the model proposed in [10] is used to account for this parasitic filtering for power spectral frequencies up to 90 kHz, for various laser wavelengths. The position detection system consists of a tunable laser used in conjunction with a Si-PIN diode and $\Delta$-$\Sigma$ data acquisition electronics with sampling frequency 195 kHz. The signal analyzed is the position of a micro-sphere doing Brownian motion in a liquid while being held in an optical trap. We thus extend the useful power-spectral frequency range in optical tweezers experiments to the maximum set by our data acquisition electronics. This enlarged bandwidth is of relevance to experiments in, e.g., single molecule biophysics [13] and microrheology [8, 14–16], as was recently demonstrated in [15] for a smaller frequency range. In the larger frequency range recorded here, parasitic filtering is stronger, so its modelling is more demanding, but not complicated, as we show. Thus the calibration procedure demonstrated here makes optical tweezers a tool of accuracy and precision with significantly higher bandwidth than before.

The paper is organized as follows: Section II briefly describes the experimental procedure. Section III gives the necessary formulas from physics and power spectrum analysis. Section IV analyzes the physics of the detection system. Section V explains how the model parameters of the photodiode detection system are determined experimentally. Sections VI and VII describe our data analysis and experimental results, respectively. Section VIII contains our conclusions.

---
[*]Present address: Department of Physics, Technical University of Denmark, DK-2800 Kgs. Lyngby, Denmark
[†]The Niels Bohr Institute, Blegdamsvej 17, DK-2100 Copenhagen Ø, Denmark
[‡]Present address: III. Physikalisches Institut, Georg-August-Universität, Friedrich-Hund-Platz 1, 27077 Göttingen, Germany

## II. EXPERIMENTS

We measured the Brownian motion of optically trapped microscopic beads in water by back-focal-plane interferometry [4]. We repeated these measurements at a number of trapping laser wavelength. We used silica beads with a diameter of 900 nm (Bangs Laboratories, Fishers, IN) dispersed in water. Beads were diluted typically to a concentration of $10^{-6}$ w/v and introduced into a sample chamber made of a cover-slip and a microscope slide glued together with double-stick tape. Beads were trapped $\sim$10 $\mu$m above the cover-slip/water interface to minimize their hydrodynamic interaction with the interface.

The custom-built instrument used for the experiments is based on a continuous-wave Ti:Sapphire laser (Mira 900F with a triple-plate birefringent filter, pumped by a Verdi V10 frequency-doubled Nd:YVO$_4$ laser, Coherent Inc., Santa Clara, Ca.), tunable from 730 to 1000 nm. The collimated beam from this laser is expanded 9 times and focused to a diffraction limited spot in the sample with a high-numerical-aperture microscope objective (S Fluor, 100x, NA 1.3, Nikon Corp., Kanagawa, Japan). The detection optics of the setup consist of a high-numerical-aperture condenser (Achromat/Aplanat, NA 1.4, Nikon) to collect the trapping laser light, with the back-focal plane of the condenser imaged onto a silicon quadrant photodiode (SPOT 9-DMI, UDT, Hawthorne, CA) operated at 15 V reversed bias. We also repeated these measurements with a largely identical setup, but with a different laser, a 1064 nm laser (Nd:YVO$_4$, Compass 1064-4000 M, Coherent Inc.) [17].

The signals from the four quadrants of the diode were amplified by low-noise, high-bandwidth preamplifiers (custom built) and the distribution of light on the diode was calculated by an analog normalizing differential amplifier (custom built) [17]. The signals were then digitized with an A/D-board sampling at 195 kHz per channel (AD16 board on a ChicoPlus PC-card, Innovative Integration, Simi Valley, CA). This board is based on $\Delta$-$\Sigma$ conversion technology [18]. This implies oversampling of the signal and no external anti-aliasing filters are needed. The specified 3dB-frequency of the board is $0.496 \times f_{\rm sample}$, in our case 96.7 kHz.

Time series of approximately 8 million points were recorded using custom-written LabView software (National Instruments, Austin, TX).

## III. TRAPPED BEAD'S BROWNIAN MOTION IN LIQUID

To make the present paper self-contained with respect to key formulae, we give those here in the notation used in [11] where more details are given.

### A. The Einstein-Ornstein-Uhlenbeck theory of Brownian motion

The Einstein-Ornstein-Uhlenbeck theory describes the Brownian motion of a spherical bead trapped in a harmonic potential with three identical, uncoupled Langevin equations, one for each of the bead's cartesian coordinates $(x(t), y(t), z(t))$ [19]. For the $x$-coordinate, this equation reads

$$m\ddot{x}(t) + \gamma_0 \dot{x}(t) + \kappa x(t) = (2k_{\rm B}T\gamma_0)^{\frac{1}{2}} \eta(t) \ . \qquad (1)$$

Here, $m$ is the mass of the bead, $\gamma_0$ its friction coefficient, $-\kappa x(t)$ the harmonic force from the trap, and the random thermal forces from the surrounding liquid are modelled with the term $(2k_{\rm B}T\gamma_0)^{\frac{1}{2}}\eta(t)$, a white-noise random process with explicitly written amplitude $(2k_{\rm B}T\gamma_0)^{\frac{1}{2}}$. The stochastic process $\eta(t)$ has vanishing mean and a delta-function as auto-correlation function. Stokes law for a spherical particle gives

$$\gamma_0 = 6\pi\rho\nu R \qquad (2)$$

where $\rho\nu$ is the liquid's shear viscosity, $\rho$ the liquid's density, $\nu$ its kinematic viscosity, and $R$ the radius of the spherical particle.

As a model for Brownian motion in a liquid, the Einstein-Ornstein-Uhlenbeck theory is only valid as a low-frequency approximation. It ignores that the friction coefficient is frequency dependent, when the hydrodynamics of the surrounding liquid is taken into account. At frequencies low enough to make the Einstein-Ornstein-Uhlenbeck theory an acceptable approximation, its inertial term, $m\ddot{x}$, can be left out to an even better approximation, which leaves us with the original Einstein theory from 1905. Thus, the only use we have for the Einstein-Ornstein-Uhlenbeck theory here, is to establish the connection between Newton's Second Law, Eq. (1), and Einstein's approximate theory for Brownian motion.

### B. Power spectrum of Einstein's theory of Brownian motion

Einstein's theory results in a Lorentzian power spectrum [20, 21] for the motion,

$$P_k \equiv \langle P_k^{\rm (ex)} \rangle = \frac{D/(2\pi^2)}{f_{\rm c}^{\ 2} + f_k^2} \ . \qquad (3)$$

Here $\langle \ldots \rangle$ denotes expectation value, $P^{\rm (ex)}$ denotes experimental power spectral values, the *corner frequency* $f_{\rm c} \equiv \kappa/(2\pi\gamma_0)$ has been introduced, and Einstein's relation $D = k_{\rm B}T/\gamma_0$ between diffusion constant, Boltzmann energy, and friction coefficient, has been used. The discrete frequency $f_k \equiv k/t_{\rm msr}$ with $k$ integer and $t_{\rm msr}$ the duration of the time interval on which $x(t)$ is Fourier transformed. Here we have used the same normalization of the power spectrum as in [11, Eqs. (7–8)]. There, it was also shown that $P_k^{(\text{ex})}$ is exponentially distributed, with expectation value given in Eq. (3) and, as is always the case for exponential distributions, with root-mean-square-deviation equal to its mean,

$$\sigma(P_k^{(\text{ex})}) = \langle (P_k^{(\text{ex})} - P_k)^2 \rangle^{\frac{1}{2}} = P_k \ . \tag{4}$$

### C. Hydrodynamically correct power spectrum of Brownian motion in liquid

The power spectrum of classical Brownian motion in a liquid is known beyond the approximate Einstein-Ornstein-Uhlenbeck theory. It is known exactly in the limit of vanishing Reynolds number, which limit is an extremely good approximation for classical Brownian motion [11]. We use this result to analyze the experimental power spectrum in its entire frequency range. By doing so we account for the frequency-dependence of viscous friction, for the frequency-dependence of the inertial mass of entrained liquid, for the bead's inertial mass, and for the finite distance $\ell$ between the center of the bead and the surface of the experimental chamber. The frequency-dependent extra friction experienced by the bead as result of its hydrodynamical interaction with this surface is known only approximately, but to a very good approximation. This is all done by replacing the power spectrum in Eq. (3) with the expression [11, 22, 23],

$$P_{\text{hydro}}(f_k; R/\ell) = \frac{D/(2\pi^2)\frac{\text{Re}\,\gamma}{\gamma_0}}{\left(f_c + f_k \frac{\text{Im}\,\gamma}{\gamma_0} - f_k^2/f_m\right)^2 + \left(f_k \frac{\text{Re}\,\gamma}{\gamma_0}\right)^2} \tag{5}$$

where to first order in $R/\ell$,

$$\gamma(f_k, R/\ell) = \gamma_0 \left(1 + (1-i)\sqrt{\frac{f_k}{f_\nu}} - i\frac{2}{9}\frac{f_k}{f_\nu}\right) \times \tag{6}$$

$$\left(1 + \frac{9}{16}\frac{R}{\ell} \times \right.$$

$$\left. \left[1 - \frac{1-i}{3}\sqrt{\frac{f_k}{f_\nu}} + \frac{2i}{9}\frac{f_k}{f_\nu} - \frac{4}{3}\left(1 - e^{-(1-i)\frac{2\ell-R}{\delta}}\right)\right]\right).$$

In these expressions, two new characteristic frequencies have been introduced: $f_\nu$ is the frequency at which the penetration depth in the liquid of the bead's linear harmonic motion equals the radius of the bead, $f_\nu \equiv \nu/(\pi R^2) = 1.6\,\text{MHz}$. The penetration depth $\delta(f) = (\nu/\pi f)^{1/2}$ characterizes the exponential decrease of the fluid's velocity field as a function of the distance from a bead that is forced to do linear harmonic motion with frequency $f$. The other characteristic frequency is $f_m \equiv \gamma_0/(2\pi m) = 1.6\,\text{MHz}$ where $m$ is the mass of the bead. The numerical values given here for these frequencies are for silica beads of $R = 450\,\text{nm}$ in water, and the two numerical values are equal because $f_m/f_\nu = 9\rho/(4\rho_{\text{bead}})$ and $\rho/\rho_{\text{bead}} \simeq 4/9$ in the present case of a silica bead in water.

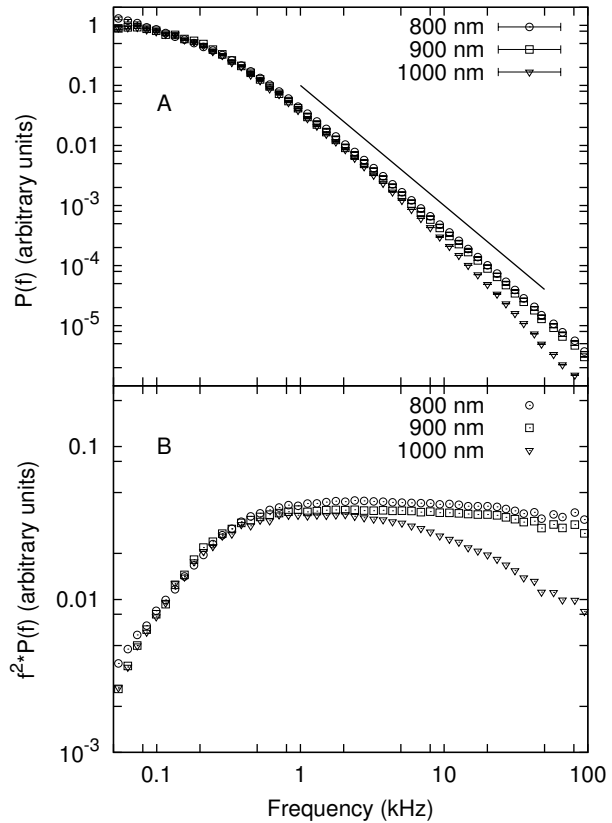

FIG. 1: Three experimental power spectra, taken with roughly equal trap strengths, using three different laser wavelengths, 800 nm, 900 nm, and 1000 nm. Panel A: Power spectrum $P(f)$ versus frequency $f$. The solid line corresponds to $P(f) \propto f^{-2}$, the behavior of a Lorentzian at large frequencies, and serves to guide the eye. Above approximately 5 kHz, the power spectrum taken with the 1000 nm laser drops off *faster* than the other two spectra, though it describes the same physical phenomenon. This faster drop-off is similar to the effect of a low-order low-pass filter. Panel B: Same data plotted as $f^2 P(f)$ versus frequency $f$ to better display the filter effect. The laser wavelength-dependence of the filter effect shows that its cannot be explained by the physics of Brownian motion. The physics of the position detection system is responsible.

This theory of trapped Brownian motion does not depend on the wavelength $\lambda$ of the trapping laser. Nevertheless, the recorded power spectra depend on $\lambda$, as illustrated in Fig. 1.

Equation (5) describes the *physical* power spectrum. A fit of it to data obtained with $\lambda = 900\,\text{nm}$ is plotted in Fig. 2. As demonstrated in Fig. 2's Panel B, the physical power spectrum differs from the *recorded* power spectrum. This discrepancy is caused by the position detection system, which low-pass filters the position signal, as described below.

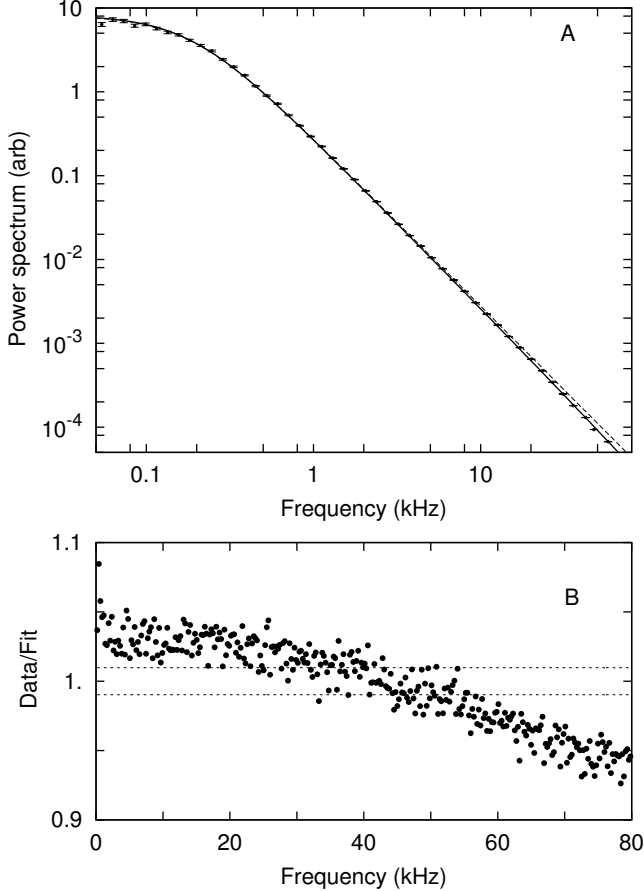

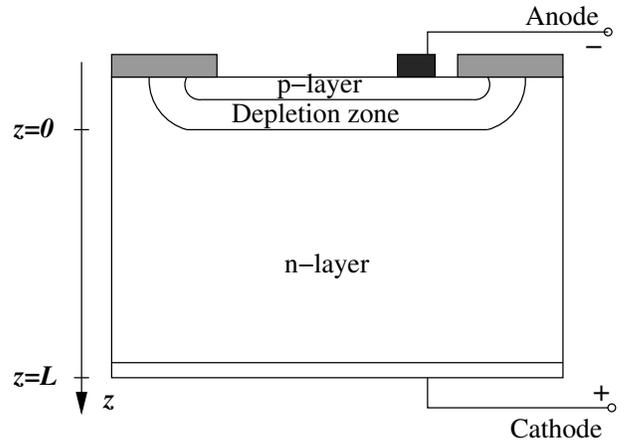

FIG. 2: Experimental power spectrum, obtained with a laser wavelength of 900 nm, fitted with theoretical power spectrum $P_{\text{hydro}}$ given in Eq. (5) in the frequency interval $[110\,\text{Hz}, 80\,\text{kHz}]$. Panel A: Blocked data points with the fitted $P_{\text{hydro}}$ shown as solid line, and a fitted Lorentzian shown as dashed line. Panel B: Residual plot—i.e., plot of experimental values divided by fitted theoretical values, $P^{(\text{ex})}(f)/P_{\text{hydro}}(f)$, versus frequency $f$. This plot reveals a significant difference between the experimental spectrum and the theory fitted to it, a difference that is not visible in Panel A. The two horizontal dashed lines indicate $\pm 1$ standard deviation, according to the theory for how the data, shown as dots, should scatter about their expectation value 1.

## IV. THE SI-PIN DIODE

Light detection with a photodiode can be modelled as follows [10]: A typical construction of a Si-PIN photodiode is illustrated in Fig. 3. The p- and n-layers are essentially field free, while there is an electrical field across the depletion zone. The nanosecond response time of the photodiode given by its manufacturer is reached when all photons are absorbed in the depletion zone, a few tens of microns into the material. This is the case for visible light. Light with longer wavelengths, however, has photon energies near or below the band gap in silicon, hence a much lower absorption coefficient. It penetrates into the substrate of the diode (commonly an n-layer), creates charge carriers also there, and these reach the depletion layer by diffusion only, thereby causing a delayed signal. (Only holes are taken into account to model the delayed signal, as argued in [10].)

The diode is very flat, its $xy$-dimensions being much larger than its $z$-dimension. We need only consider diffusive motion and only in the $z$-direction: Significant concentration gradients of holes, hence significant transport of these by diffusion, are found only in the $z$-direction, because holes are removed only in the depletion zone. Recombination of holes with conduction electrons can be ignored to a very good approximation on the time scales and with the level of precision considered here.

FIG. 3: Typical cross-section of a photodiode operated with reverse bias. Reproduced from [28]. Not to scale. While the sensitive area measures centimeters square, $L$ is typically some hundred micrometers, and the p-layer and the depletion zone are typically a few to tens of micrometers thick.

### A. Diffusion equation

The time evolution of the density $\rho(z,t)$ of holes as function of depth $z$ within the substrate is governed by the diffusion equation

$$\frac{\partial \rho}{\partial t} = \mathcal{D}\frac{\partial^2 \rho}{\partial z^2} \qquad (7)$$

where $\mathcal{D}$ is the thermal diffusion coefficient of holes. In pure silicon, $\mathcal{D} = 12.9\,\text{cm}^2/\text{s}$ at room temperature [24, Fig. 4.11], [6, Example 5.1]. In the weakly doped n-layer that makes up the substrate, $\mathcal{D}$'s value in pure silicon is a good approximation, and this value will be used in the following.

As shown in Fig. 3, we choose the $z$-axis so that the n-layer's boundaries are located at $z = 0$ and $z = L$. They are modelled as absorbing at $z = 0$ and reflecting at $z = L$, i.e.,

$$\rho(z = 0, t) = 0 \; , \quad \frac{\partial \rho}{\partial z}(z = L, t) = 0 \; . \tag{8}$$

Holes are detected within nanoseconds once they have reached the depletion zone. The delayed part of the diode's output signal is therefore well approximated by the diffusive current of holes out of the n-layer and into the depletion layer. This current, $I(t)$, is given by Fick's law,

$$I(t) = -\mathcal{D}\frac{\partial \rho}{\partial z}(z = 0, t) \; . \tag{9}$$

The solution to the diffusion equation, Eq. (7), in any compact volume can be written as a discrete sum of eigenmodes of exponential relaxation, each with its own relaxation time. With the boundary conditions of Eq. (8), Eq. (7) gives

$$\rho(z, t) = \rho_0 \sum_{n=0}^{\infty} b_n \exp(-t/\tau_n) \sin\left(\frac{(2n+1)\pi}{2} \frac{z}{L}\right) \; , \tag{10}$$

where the characteristic relaxation time $\tau_n$ of the $n$th spatial mode is

$$\tau_n \equiv \mathcal{D}^{-1}\left(\frac{2L}{(2n+1)\pi}\right)^2 \; . \tag{11}$$

The relaxation times decrease rapidly with $n$, the longest being

$$\tau_0 = \frac{4L^2}{\pi^2 \mathcal{D}} \; , \tag{12}$$

with the higher spatial modes having shorter relaxation times by factors 9, 25, 49, 81,....

This large separation between the slowest relaxation mode and the faster ones gives rise to another significant simplification. In the case considered in [10], the maximum frequency considered was as low as 14 kHz. There all relaxation modes except the slowest one were too fast to be resolved in time and instead contributed to the part of the output signal that appeared instantaneous. Also, silicon is so transparent to the 1064 nm laser light used that the initial distribution of charge carriers created by a flash of light was considered constant in the $z$-direction. Specifically, the inverse absorption coefficient of 1064 nm light in Si is of order 1 mm whereas the dimension of a typical diode in the $z$-direction is of order a few hundred $\mu$m. These simplifying circumstances made it possible to use a simple solution of the model.

### B. Solution of the diffusion equation for a large range of wavelengths

Here, we discuss a range of wavelengths of incident laser light and aim to model the experimental power spectrum up to 100 kHz. The density of holes created by a flash of light then takes the form $\rho(z, t = 0) = \rho_0 \exp(-az)$ where $a$ is the absorption coefficient of silicon at the particular wavelength of light. With Eqs. (7–12) we thus find

$$b_n = 2\frac{(-1)^{(n+1)}(aL)\exp(-aL) + \frac{(2n+1)\pi}{2}}{(aL)^2 + \left(\frac{(2n+1)\pi}{2}\right)^2} \tag{13}$$

which reduces to the result of Ref. [10] when $a$ vanishes. The diffusive current of holes $I(t)$ out of the n-layer is

$$I(t) = \frac{\pi \mathcal{D} \rho_0}{2L} \sum_{n=0}^{\infty} (2n+1) b_n \exp(-t/\tau_n) \; . \tag{14}$$

For a sufficiently small absorption coefficient, and with a Nyquist frequency $f_{\text{Nyq}} = 97.5$ kHz as here, several relaxation modes may be discerned in the data. Therefore, we model the response function $g(t)$ of the photodiode with a yet undetermined number of terms,

$$g(t) = \alpha^{(\text{diode}, N)} \delta(t) + \tag{15}$$
$$(1 - \alpha^{(\text{diode}, N)}) C_N \sum_{n=0}^{N} (2n+1) b_n \exp\left(-\frac{t}{\tau_n}\right) \; ,$$

where $t$ is the duration of time from the moment a pulse of light hits the diode till an output current is detected. The fraction of response that effectively is instantaneous when $N$ relaxation modes are included in the sum, is denoted $\alpha^{(\text{diode}, N)}$. The factor $C_N$ is a normalization factor making the time integral of the second term in $g(t)$ equal to $1 - \alpha^{(\text{diode}, N)}$, i.e., the fraction of the response that is *not* effectively instantaneous. Thus

$$C_N = \left(\sum_{n=0}^{N} (2n+1) b_n \tau_n\right)^{-1} \equiv \left(\sum_{n=0}^{N} \zeta_n\right)^{-1} \; . \tag{16}$$

We see that a pulse of light will, in principle, cause output at all later times due to the exponentials in Eq. (15), but we also see that these currents die out exponentially fast, with characteristic relaxation times given in Eq. (11).

A light signal $S(t)$ detected by the photodiode thus produces an output

$$S^{(\text{del})}(t) = \int_{-\infty}^{t} g(t-t') S(t') \mathrm{d}t' \; , \tag{17}$$

part of which is delayed. In Fourier space, by virtue of the convolution theorem, this relationship reads

$$\widetilde{S}^{(\text{del})}(f) = \widetilde{S}(f) \cdot \tilde{g}(f) \; , \tag{18}$$

where $\sim$ denotes Fourier transformation. Thus, the *recorded* experimental power spectrum $P^{(\text{ex})}(f) \equiv \left\langle |\widetilde{S}^{(\text{del})}(f)|^2 \right\rangle$ is simply the power spectrum of the physical signal $\left\langle |\widetilde{S}(f)|^2 \right\rangle$ multiplied with $G(f) \equiv |\tilde{g}(f)|^2$.

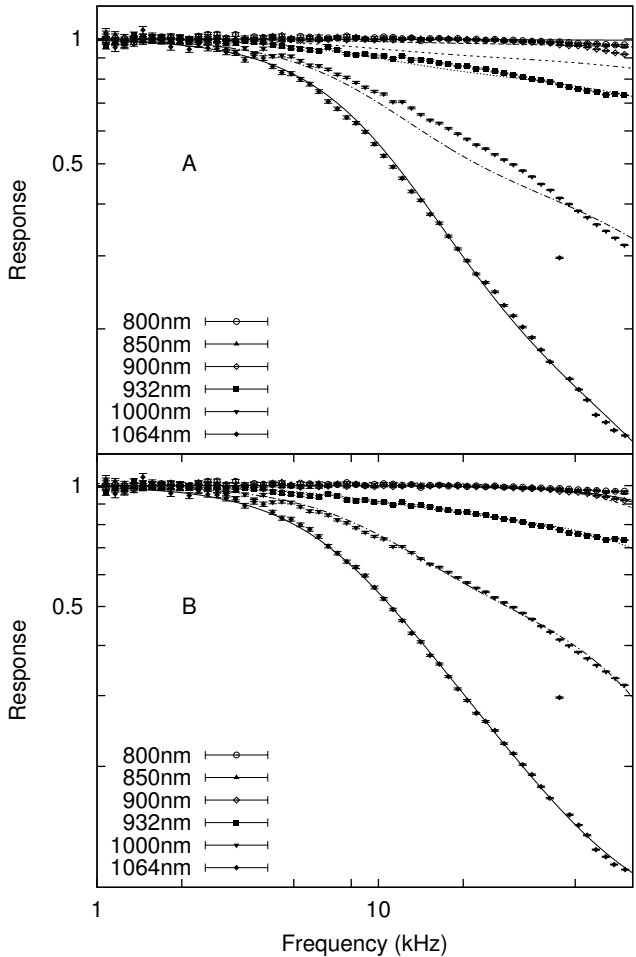

FIG. 4: Comparison of experimental data for the characteristic attenuation function $G(f)$ with theoretical models for $G(f)$. Systematic deviations of data points from the value 1 reflect the low-pass filtering caused by the photo diode. The legend gives the correspondence between symbols and laser wavelengths. The theoretical attenuation function does not differ visibly from the constant 1 for $\lambda \leq 850$ nm. **A**: Insufficiency of model given in Eqs. (19), (11), and (13). *Points*: Blocked experimental spectra $P^{(ex)}(f)$ for various laser wavelengths, divided by $P_{hydro}(f)$. The parameters in $P_{hydro}(f)$ were determined from a fit to the spectra at low frequencies, up to $f_{max} = 2$ or 5 kHz. *Lines*: $G(f)$ of Eq. (19) with $N=5$. Parameter values were obtained by fitting $G(f)$ to 1064 nm data (with the outlier near 40 kHz excluded. This noise-peak is caused by some source in the building that we could not get rid of.). This results in $f_0 = 9.8 \pm 0.1$ kHz, $L = 0.80 \pm 0.17$ mm, and $L^{(dep)} = 68 \pm 3\,\mu$m. The values for $L$ and $L^{(dep)}$ correspond to $\alpha^{(diode)} = 0.12$ for a wavelength of $\lambda = 1064$ nm. This fit is shown as the solid line through the 1064 nm data. Other lines use the same values for $f_0$, $L$, and $L^{(dep)}$. Upper solid line: $\lambda = 800$ nm, long-dashed: $\lambda = 850$ nm, short-dashed: $\lambda = 900$ nm, dotted: $\lambda = 932$ nm, dot-dashed: $\lambda = 1000$ nm. **B**: Sufficiency of model given in Eqs. (19), when used as phenomenological model *without* Eqs. (11) and (13). Only two slowest relaxation modes are needed, i.e., $N=1$. Same experimental data as in A.

Upon Fourier transformation of Eq. (15), we find

$$
\begin{aligned}
G(f) = {} & (\alpha^{(diode,N)})^2 + \\
& (1-\alpha^{(diode,N)})^2 C_N^2 \left\{ \sum_{n=0}^{N} \frac{(\zeta_n)^2}{1+(f/f_n)^2} + \right. \\
& \left. 2 \sum_{n=0}^{N} \sum_{n'<n} \frac{\zeta_n \zeta_{n'} (1+(f/f_n)(f/f_{n'}))}{(1+(f/f_n)^2)(1+(f/f_{n'})^2)} \right\} \\
& + 2\alpha^{(diode,N)}(1-\alpha^{(diode,N)}) C_N \sum_{n=0}^{N} \frac{\zeta_n}{1+(f/f_n)^2} .
\end{aligned}
\quad (19)
$$

In this expression, we have introduced $f_n \equiv 1/(2\pi\tau_n) = (2n+1)^2 f_0$. In the case $N=0$, Eq. (19) reduces to the result given in [10],

$$
G_0(f) = \alpha^{(diode,0)2} + \frac{1-\alpha^{(diode,0)2}}{1+(f/f_0)^2} \quad (20)
$$

where $f_0$ is the same as $f_{3dB}^{(diode)}$ in [10].

The theoretical curves plotted in Fig. 4A are $G(f)$ in Eq. (19) with parameters $L$, $L^{(dep)}$ and $f_0$ determined by fitting $G(f)$ to the data taken with $\lambda=1064$ nm, and using these values also when comparing $G(f)$ to data taken at different wavelengths, while varying only the absorption coefficient $a$. The curves approximately follow the data. This demonstrates that our simple physical model goes a long way in describing what goes on, but also that we need to refine it a little in order for it to be useful for precision calibration.

In the version discussed till now, we have assumed a sharp boundary between a field-free n-layer and a field-filled depletion-layer. This is not an exact rendition of reality, but a good approximation as long as one does not probe the spatial structure of the diode with a resolution fine enough to resolve the gradual transition between the n-layer and the depletion zone. When one increases $f_{max}$, higher relaxation modes for the diffusion equation are needed to describe the photo diode's response function. The spatial components of these relaxation modes are eigenfunctions for the Laplacian operator, with a number of nodes equal to the index $n$. Thus, as we increase $f_{max}$, and with it our experimental time resolution of the diode's response function $g(t)$, we consequently increase also the resolution with which the experiment probes the internal spatial geometry of the diode.

This calls for less ideal assumptions than those used above. We replace the simple one-dimensional diffusion equation Eq. (7) by a diffusion equation in three dimensions. We still treat the depletion zone as an absorbing boundary on the region that we refer to as the n-layer. We also still treat the n-layer as field-free, so no convective term occurs in the diffusion equation. Doing this, and using the fact that the n-layer is a compact volume, we need not solve the more complicated equation. We know from mathematical spectral analysis that the solu-

tion for the density of holes can be written

$$\rho(x,y,z,t) = \rho_0 \sum_{n=0}^{\infty} b_n \exp(-t/\tau_n) h_n(x,y,z) ,  \quad (21)$$

where the functions $h_n(x,y,z)$ are unknown spatial eigenfunctions of the diffusion equation in the n-layer. The values of $b_n$ are restricted by normalization, but apart form this, $b_n$ and $b_{n'}$, as well as $\tau_n < \tau_{n'}$, are now treated as unrelated for $n < n'$. Thus, the response of the diode retains the structure of Eqs. (15) and (19) with a discrete spectrum of well-separated relaxation times, but the specific expressions in Eqs. (11) and (13) no longer hold.

Consequently, the theoretical expression, $G(f)P_{\text{hydro}}(f)$, for the recorded spectrum remains valid, but the parameters $f_n$ and $\zeta_n$ in it are unknown a priori, and their values must be chosen by fitting $G(f)P_{\text{hydro}}(f)$ to the experimental spectrum, while respecting the normalization condition of $\zeta_n$. Fig. 4B demonstrates that with the inclusion of only two modes in the filtering function, we are able to fit the experimental spectra very well.

## V.  DIODE PARAMETERS

Below, we fit the theoretical model, $G(f)P_{\text{hydro}}(f)$ to recorded experimental spectra with $f_0$ and $\alpha^{(\text{diode})}$ as fitting parameters. These parameters are related to the dimensions of the diode as follows: From Eq. (12) we have

$$f_0 = \frac{\pi \mathcal{D}}{8L^2} . \quad (22)$$

Since neither the diffusion coefficient for holes, $\mathcal{D}$, nor the depth of the n-layer, $L$, depends on the laser wavelength, we expect fits of our model to result in values for $f_0$ that are independent of the laser wavelength used to obtain the data fitted to. The parameter $\alpha^{(\text{diode})}$ denotes the fraction of light that is detected "instantaneously" according to the physical model.

With the coordinates shown in Fig. 3, the number of holes created in the depletion layer is

$$\int_{-L^{(\text{dep})}}^{0} \rho_0 \exp(-az) \mathrm{d}z = \frac{\rho_0}{a} \left( \exp(aL^{(\text{dep})}) - 1 \right) , \quad (23)$$

where $L^{(\text{dep})}$ is the thickness of the depletion layer. $L^{(\text{dep})}$ depends on the bias voltage applied across the photodiode and it is not available from the manufacturer of the diode. The number of charge carriers in the slowest $N+1$ relaxation modes is

$$\sum_{n=0}^{N} \int_0^L b_n \sin\left(\frac{(2n+1)\pi}{2} \frac{z}{L}\right) \mathrm{d}z$$

$$= \sum_{n=0}^{N} b_n \frac{2L}{(2n+1)\pi} . \quad (24)$$

The total number of holes created is

$$\int_{-L^{(\text{dep})}}^{L} \rho_0 \exp(-az) \mathrm{d}z = \frac{\rho_0}{a} \left( \exp(aL^{(\text{dep})}) - \exp(-aL) \right) . \quad (25)$$

Thus, $\alpha^{(\text{diode})}$ can be determined: From the results of Eqs. (23–25), the fraction of light detected instantaneously is

$$\alpha^{(\text{diode},N)} = \frac{\left(\exp(aL^{(\text{dep})}) - 1\right) + \sum_{n=N+1}^{\infty} \frac{ab_n}{\rho_0} \frac{2L}{(2n+1)\pi}}{\left(\exp(aL^{(\text{dep})}) - \exp(-aL)\right)}$$

$$\simeq \frac{\left(1 - \exp(-aL^{(\text{dep})})\right)}{\left(1 - \exp(-a(L + L^{(\text{dep})}))\right)} \quad (26)$$

where the last result holds for a sufficiently large value of $N$.

## VI.  DATA ANALYSIS

Power spectra for recorded positions of the bead were calculated without use of so-called *windowing* [25]. Instead, noise reduction was achieved by *blocking* [11, 25, 26], i.e., blockwise averaging. The recorded power spectra were then fitted with the model spectra: In [11], the recorded experimental spectrum was modelled with the filtered and aliased physical spectrum. The physical spectrum is $P_{\text{hydro}}(f; R/\ell)$ given in Eq. (5), with $R = 450$ nm in our case, and $\ell = 10\,\mu$m. The data analyzed in the present paper were recorded with a data acquisition system that uses $\Delta$-$\Sigma$ technology with 64 times over-sampling in its ADC [18]. Consequently, there is no discernible aliasing to account for in the experimental spectrum, and we didn't. Filtering by the photodiode detection system we did account for, as described in Eq. (15). The functional form of any additional electronic filtering was unknown. So we looked for additional filters in the electronics in an independent measurement with a signal generator. We found a flat response below 80–90 kHz (see also Fig. 7A), and we consequently restricted our analysis of the power spectrum to frequencies below 80 kHz. We assumed no filtering of the physical spectrum took place below this frequency, except the parasitic one in the photo diode. Combined with the absence of aliasing, this means that the recorded spectrum can be compared directly with the physical spectrum after the latter has been filtered only by the parasitic filter in the photodiode.

All data series analyzed were consistent with a hookean force, as assumed in the theory with which we analyze the data. An example is shown in Fig. 5.

## VII.  RESULTS

We discuss first our fitting procedure, then present the results.

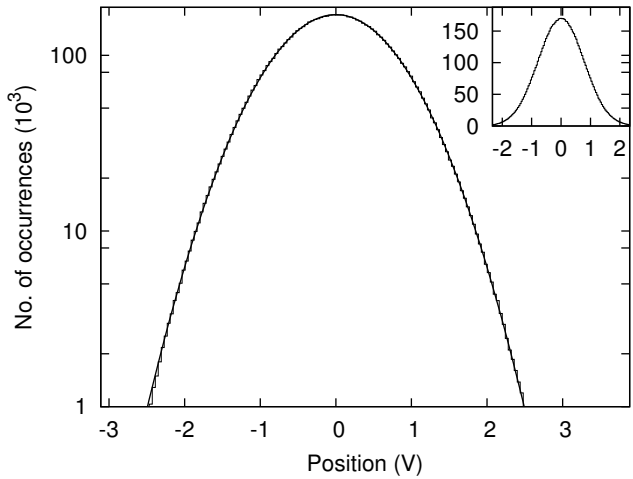

FIG. 5: Lin-log plot of histogram of values visited by one coordinate in the position time-series recorded with laser wavelength 900 nm (same data as in Fig. 2), overlayed with a Gaussian of same width and height. The inset shows the same histogram in a lin-lin plot.

## A. Defining a good fit

All spectra were least-square fitted with either $P_{\rm hydro}(f)$ of Eq. (5) or with $G(f)P_{\rm hydro}(f)$, with $G(f)$ given in (19). To judge whether a fit was acceptable or not, the support/goodness-of-fit [25, 27] was evaluated. The goodness-of-fit will not reveal small *systematic* discrepancies between theory and data as long as $\chi^2$ per degree of freedom is near 1. To be able to detect such errors, we also construct "residual plots" by dividing data values by fitted values, cf. Figs. 2B, 6A, and 7A.

In least-squares fitting, each data point enters the expression for $\chi^2$ with a weight factor that depends on the error bar on that point. More correctly, the weight factor depends on the standard deviation of the Gaussian distribution with respect to which the data points scatter about the expectation value, the fitted function. This standard deviation is usually estimated by the experimental error bar on the data point. We know its value, however, hence need not estimate it. The theory that gives the expectation value, gives the standard deviation as well, with no additional fitting parameters introduced [11], but with a small addition to $\chi^2$ so the quantity minimized is $\bar{\mathcal{F}}_2$ in [11, Eq. (E.7)]. In the following, we accept or do not accept fits depending on their support and their residual plots, and use the information from these fits to draw conclusions about the model.

## B. Results of fits

When fitting $P_{\rm hydro}(f)$ or $G(f)P_{\rm hydro}(f)$ to experimental data, we have the option to vary both the maximum frequency fitted to, $f_{\rm max}$, and, for $G(f)P_{\rm hydro}(f)$, also the number $N+1$ of relaxation modes included in the description of the response function of the diode.

### 1. Theory without parasitic filter.

As apparent already from Fig. 2B, even for the shortest wavelengths investigated, no acceptable fit of $P_{\rm hydro}$ was possible with $f_{\rm max} = 80\,\rm kHz$. Table I summarizes our results for fits done without accounting for the diode's parasitic filter. We see that for frequencies up to at least 40 kHz, $P_{\rm hydro}(f)$ describes the recorded spectrum for wavelengths up to $\lambda = 900\,\rm nm$. So in this range the detection system is not an unintended filter.

TABLE I: Laser wave lengths and power spectrum frequencies for which the position detection system is *not* a parasitic filter. Values for $f_{\rm max}$ for which fits of $P_{\rm hydro}$ to the experimental power spectra have at least 1% support. Tested values of $f_{\rm max}$ were 2 kHz, 5 kHz, 10 kHz, 15 kHz, 25 kHz, 40 kHz, 60 kHz and 80 kHz.

| $\lambda$ (nm) | $f_{\rm max}$ (kHz) |
|---|---|
| 750 | <60 |
| 800 | <60 |
| 850 | <60 |
| 900 | <60 |
| 915 | <25 |
| 932 | <10 |
| 944 | <5 |
| 962 | <5 |
| 984 | <5 |
| 1000 | <5 |

### 2. Theory with parasitic filter with only the slowest diffusion mode included.

With only the slowest relaxation mode included in the description of parasitic filtering by the diode—e.g., with $G(f) = G_0(f)$ of Eq. (20)—we obtained fits with a support of at least 1% for $f_{\rm max} = 80\,\rm kHz$ and for wavelengths up to $\lambda = 915\,\rm nm$. An example is shown in Fig. 6. The quality of the fit is demonstrated both through the support of 82%, and visually by the residual plot shown in panel A and the perfect exponential distribution shown in panel C, cf. [11]. Figure 2 shows the result of an attempt to fit the same data with $P_{\rm hydro}$ alone up to $f_{\rm max} = 80\,\rm kHz$.

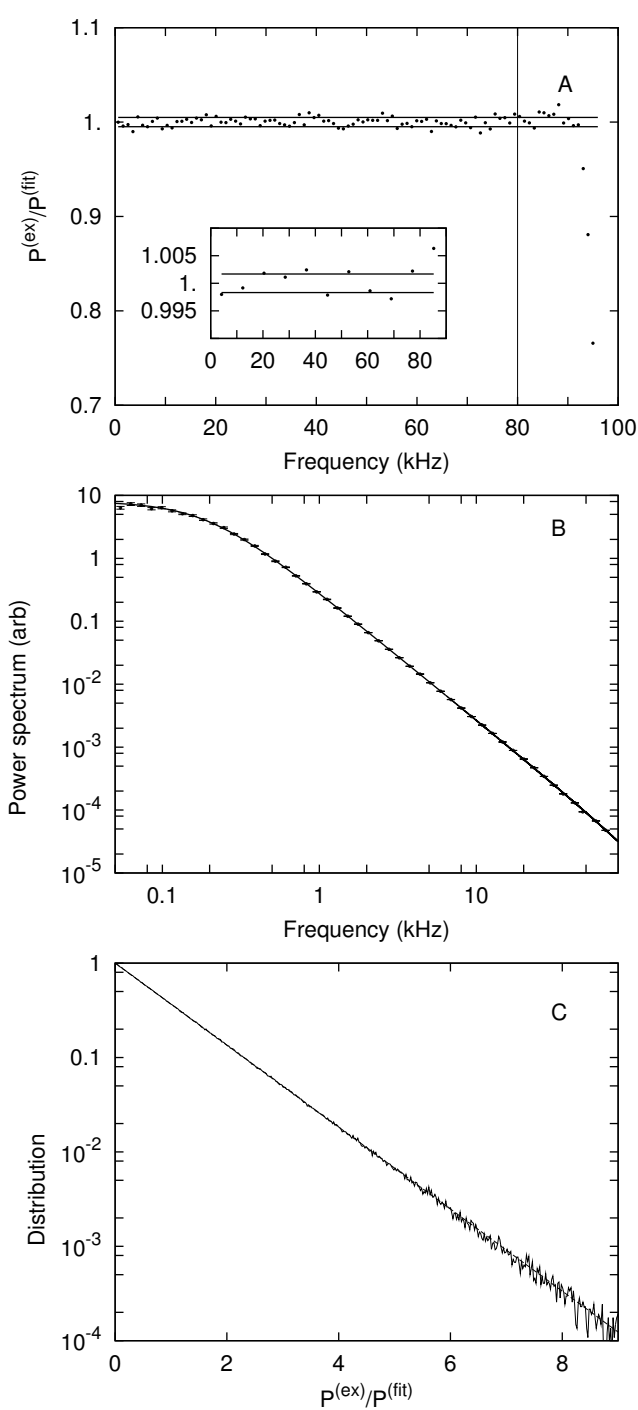

FIG. 6: Experimental power spectrum obtained with laser wavelength 900 nm (same data as in Fig. 2) and fitted with one diffusion mode in the parasitic filter, i.e., $G(f) = G_0(f)$ in Eq. (20). The frequency range of the fit was [110 Hz, 80 kHz]. A: Residual plot of data/fit. The two horizontal lines show ± 1 standard deviation, known from the theory. The backing of the fit is 82%. Inset: Same residual plot of data/fit after further block-averaging of data. Two horizontal lines show ± one standard deviation according to theory, here ±1.7 per mil. The data points scatter a little more than ideally expected for normally distributed data. This may be because the data acquisition board is slightly non-linear, with ripples in its characteristic function of max ±0.005 dB amplitude, i.e., ±1.3 per mil. We have insufficient precision to resolve such ripples in our power spectrum of Brownian motion, but we *have* reached the limit on achievable precision set by these ripples. B: Experimental power spectrum (data points with error bars) and fitted theory with parasitic filter $G_0(f)$ (solid line). C: Histogram of $N = 3 \cdot 10^6$ experimental power spectral values $P^{(ex)}(f_k)$, measured in units of their expectation values $P^{(fit)}(f_k)$, the latter being the fit shown in Panel B. Dashed line: $\exp(-x)$, the distribution

*3. Theory with parasitic filter with all diffusion modes, related by Eqs. (11) and (13).*

For wavelengths larger than 915 nm, we investigated the model assuming an infinitely sharp boundary between depletion layer and n-layer, i.e., with $G(f)$ defined through Eqs. (19), (11), and (13). We varied the value of $f_{\max}$ and the number of modes included in $G(f)$. With $f_{\max} > 20$–30 kHz, and for wavelengths larger than ~950 nm, no acceptable fits were found. Residual plots like Fig. 2, showed systematic deviations from the theory.

*4. Theory with phenomenological parasitic filter: Two slowest modes are sufficient.*

We consequently dropped the assumption leading to Eqs. (11) and (13). Instead, we maintain the discrete spectrum of relaxation times that is characteristic for diffusion out of any compact region, but treat each relaxation time and its relative importance as fitting parameters. As we shall see, physically realistic values result from fitting these parameters. The usefulness of the model depends on more than its realism: It also depends on whether the model makes so much more of the power spectrum meaningful and interpretable that it is worth the cost of the model's extra parameters. It does, as we shall see.

We found that with $f_{\max} = 80$ kHz, fits with at least 1% support can be obtained for the entire range of laser wavelengths available, and with the inclusion of only 2 relaxation modes in the description of the diode. Inclusion of more modes did not improve the quality of the fits which is consistent with $f_0 \sim 10$ kHz → $f_1 \sim 90$ kHz and $f_2 \sim 250$ kHz. Each extra mode adds two more parameters to be fitted, $f_n$ and $\zeta_n$ in Eq. (19), and all parameters were determined only with larger errors for a fit of otherwise similar quality. Therefore, only fits with 2 relaxation modes are considered below. Figure 7 shows a fit to data obtained with the 1064 nm laser. This is the longest laser wavelength considered here, hence the case of lowest absorption coefficient in the detector. The data consequently demonstrate the strongest parasitic filtering considered here, and therefore demand the most from our model for the phenomenon.

From this and similar fits, we determined values for the parameters that describe the diode. These values are plotted in Fig. 8. The trends in these data agree with expectations based on known absorption characteristics of Si: In Fig. 8AB, the solid curves show $\alpha^{(\text{diode})}$ and $\zeta_0/(\zeta_0 + \zeta_1)$ according to Eqs. (10–13). The values used for $L^{(\text{dep})}$ and $L$ are fitted, since their exact values are proprietary information. A total thickness of the diode of order 600 μm is, to our knowledge, reasonable for typical diode-dimensions. Panels C and D of the figure show that at wavelengths where the values of $f_0$ and $f_1$ are well determined—i.e, where parasitic filtering is significant—their values are roughly independent of the wavelength,

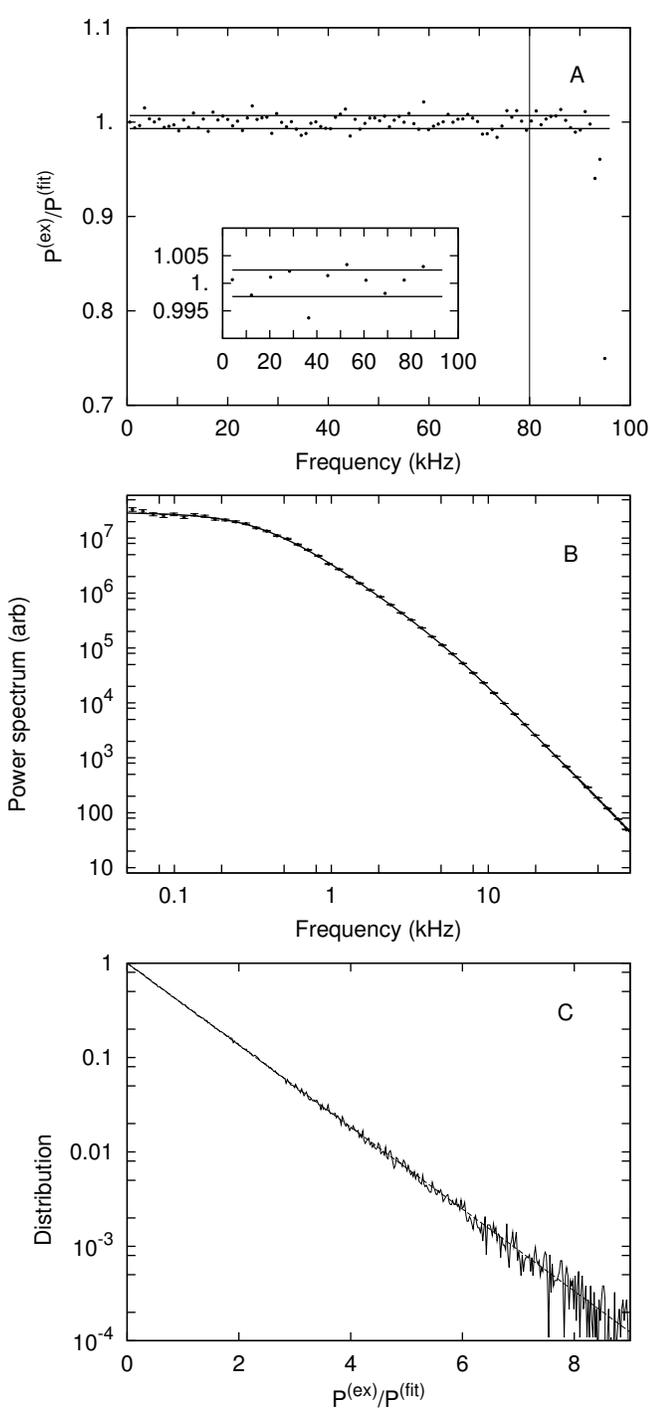
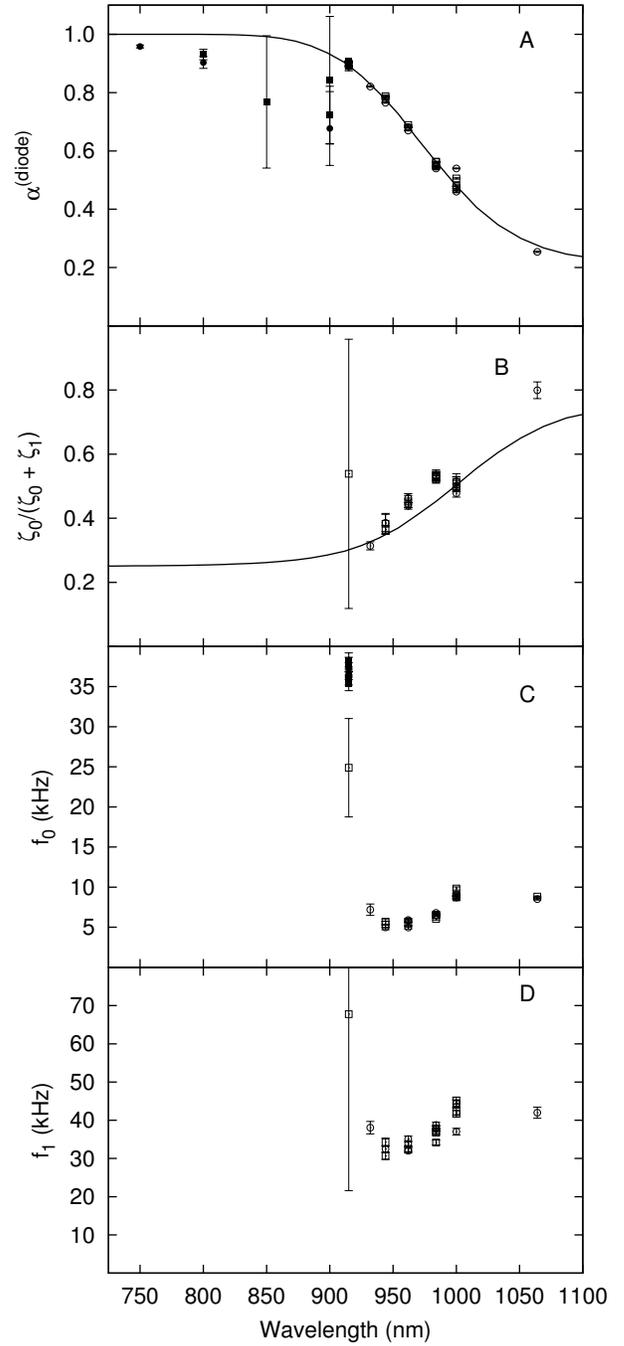

FIG. 7: Experimental power spectrum obtained with 1064 nm laser and fitted with $P^{(\text{fit})}(f) = G(f)P_{\text{hydro}}(f)$, using the phenomenological version of $G(f)$ in Eq. (19) *without* Eqs. (11) and (13), and with just two diffusion modes, $N = 1$. The value of $\chi^2$ per degree of freedom is 1.023, resulting in a backing of 10%. The maximum frequency fitted to was $f_{\max} = 80$ kHz. A: Residual plot. All data are shown, but only data of frequencies below 80 kHz (straight vertical line) were fitted to. The dots are $P^{(\text{ex})}/P^{(\text{fit})}$. Two horizontal lines show ±1 standard deviation, known from the theory for the power spectrum. Note that the data points seem filtered with a roll-off frequency near 90 kHz. Inset: Same residual plot of data/fit after further block-averaging of data. Two horizontal lines show ± one standard deviation according to theory, here ±2.4 per mil. B: Power spectrum versus frequency in a log-log plot. Data points with error bars, most of which are too small to be seen. The solid line is the fit. C: Histogram of $N = 8 \cdot 10^5$ experimental power spectral values $P^{(\text{ex})}(f_k)$, measured in units of their expectation values $P^{(\text{fit})}(f_k)$, the latter being the fit shown in panel B. Dashed line: $\exp(-x)$, the theoretical distribution for $P^{(\text{ex})}(f_k)/P^{(\text{fit})}(f_k)$.

FIG. 8: Fitting parameters that describe the photodiode, obtained from fits with at least 1% support, shown as a function of wavelength. Squares: $x$-coordinate. Circles: $y$-coordinate. Open symbols: two diffusional modes in diode-filter. Filled symbols: one diffusional mode in diode-filter. A: The relative amount $\alpha^{(\text{diode})}$ of the position signal that is detected instantly, as function of laser wavelength. The solid line shows a fit of the theory in Eq. (26). It resulted in $L^{(\text{dep})} = 128 \pm 1$ $\mu$m and $L = 458 \pm 31$ $\mu$m. B: The relative importance of the slowest relaxation mode in the parasitic filter, shown as its relative amplitude $\zeta_0/(\zeta_0 + \zeta_1)$. The solid line shows this ratio for the theory fitted as in Panel A. C and D: Frequencies of the $0^{\text{th}}$ and $1^{\text{st}}$ mode in the diode filter, resulting from fits at various wavelengths. These values are approximately independent of wavelength, as expected. Note the higher value of $f_0$ for 915 nm, obtained in fits with only one mode included.

as expected from our physical model of the phenomenon.

## VIII. CONCLUSION

When an infrared laser and a common silicon photodiode are used together in a position detection system, the photodiode is also an unintended low-pass filter. With 1064 nm light, 90% of the signal power is lost at 80 kHz; with 900 nm light, 10% of the signal power is lost at 80 kHz; see Fig. 4. The physics of this "parasitic filtering" phenomenon is understood, and is described by simple one-dimensional diffusion out of a 1D box with an absorbing boundary. The physical dimensions of the problem are such that only one or two exponential relaxation modes are needed in order to describe the filter effect accurately up to 160 kHz sampling frequency; see Figs. 6 and 7: The 1% stochastic scatter in the experimental power spectral values shown there is *too large* for us to resolve any inadequacies in our description of the parasitic filter's characteristic function! Thus the filter characteristics in Eq. (20) with $N = 0$ or 1 is all it takes to extend the useful bandwidth of this type of position detection system, typical for optical tweezers, from as low as 1 kHz and up to 80 kHz and possibly beyond. The upper limit in frequency encountered in the present paper is not set by our description, but by the electronics used.

A number of alternative photodetectors, of different material or construction, have been applied over the last couple of years [5, 12] in optical tweezers systems where infrared lasers are used for position detection. Such detectors provide a large increase in the useful bandwidth, independent of the procedure of analysis. Interestingly, however, and currently unexplained, signs of parasitic filtering with a 3dB frequency of order 70 kHz have also been observed in experiments where 1064 nm light was detected by an InGaAs diode [29]. The instrumental (electronic) bandwidth of detection should not be confused with filtering intrinsic to the experiment itself, which we don't discuss here. For example, the motions of a single motor protein molecule attached to an optically trapped bead, will be low-pass filtered by the mechanical response of the bead in a viscous solvent [21].

In summary, we have demonstrated here how the electronic bandwidth of detection for even common silicon photodetectors used with near IR lasers can be increased to frequencies approaching 0.1 MHz with high precision and have provided the tools of analysis necessary to achieve this. These tools can be easily adapted to similar situations commonly encountered in modern optical trapping experiments.

### Acknowledgments

This work was funded by a VIDI-grant from the *Research Council for Earth and Life Sciences (ALW)* to EP and grants from the *Foundation for Fundamental Research on Matter (FOM)*, both with financial support from the *Netherlands Organization for Scientific Research (NWO)*. KBS was funded by the *Danish Research Councils* and the *Carlsberg Foundation*. We thank the *Colloid Synthesis Facility*, Utrecht University, for kindly providing silica beads and Meindert van Dijk for his involvement in the early stages of this project.

[1] G. Meyer and N. M. Amer. *Appl. Phys. Lett.*, 53:1045–1047, 1988.
[2] S. Alexander, L. Hellemans, o. Marti, J. Schneir, V. Elings, P. K. Hansma, M. Longmire, and J. Gurley. *J. Appl. Phys.*, 65:164–167, 1989.
[3] A. D. Mehta, J. T. Finer, and J. A. Spudich. *Meth. Cell Biol.*, 55:47–69, 1998.
[4] F. Gittes and C. F. Schmidt. *Opt. Lett.*, 23:7–9, 1998.
[5] Keir C. Neuman and Steven M. Block. *Rev. Sci. Ins.*, 75:2787–2809, 2004.
[6] Gerhard Lutz. *Semiconductor Radiation Detectors.* Springer-Verlag, Berlin Heidelberg, 1999.
[7] C. Yang, D. N. Jamieson, S. M. Hearne, C. I. Pakes, B. Rout, E. Gauja, A. J. Dzurak, and R. G. Clark. *Nucl. Instrum. Methods Phys. Res. B*, 190:212, 2002.
[8] F. Gittes, B. Schnurr, P. D. Olmsted, F. C. MacKintosh, and C. F. Schmidt. *Phys. Rev. Lett.*, 79:3286–3289, 1997.
[9] C. Veigel, M. L. Bartoo, D. C. W. White, J. C. Sparrow, and J. E. Molloy. *Biophys. J.*, 75:1424, 1998.
[10] K. Berg-Sørensen, L. Oddershede, E.-L. Florin, and H. Flyvbjerg. *J. Appl. Phys.*, 93:3167–3176, 2003.
[11] Kirstine Berg-Sørensen and Henrik Flyvbjerg. *Rev. Sci. Ins.*, 75:594–612, 2004.
[12] Erwin J. G. Peterman, Meindert van Dijk, Lukas C. Kapitein, and Christoph F. Schmidt. *Rev. Sci. Ins.*, 74:3246–3249, 2003.
[13] M. J. Lang, C. L. Asbury, J. W. Shaevitz, and S. M. Block. *Biophys. J.*, 83:491–501, 2002.
[14] E. Helfer, S. Harlepp, L. Bourdieu, J. Robert, F. C. MacKintosh, and D. Chatenay. *Phys. Rev. Lett.*, 85:457–460, 2000.
[15] Iva Marija Tolić-Nørrelykke, Emilia-Laura Munteanu, Genevieve Thon, Lene Oddershede, and Kirstine Berg-Sørensen. *Phys. Rev. Lett.*, 93:078102–1–4, 2004.
[16] M. Atakhorrami, G. H. Koenderink, C. F. Schmidt, and F. C. MacKintosh. *Phys. Rev. Lett.*, 95:208302, 2005.
[17] M. W. Allersma, F. Gittes, M. J. deCastro, R. J. Stewart, and C. F. Schmidt. *Biophys. J.*, 74:1074–1085, 1998.
[18] AD7722. Datasheet, Rev. 0, 1996.
[19] R. Kubo, M. Toda, and N. Hashitsume. *Statistical Physics*, volume 2. Springer-Verlag, Heidelberg, 1985.
[20] Karel Svoboda and Steven M. Block. *Ann. Rev. Biophys. Biomol. Struct.*, 23:247–285, 1994.
[21] Frederick Gittes and Christoph F. Schmidt. *Methods Cell Biol.*, 55:129–156, 1998.
[22] Kirstine Berg-Sørensen and Henrik Flyvbjerg. Erratum

to: *Rev. Sci. Ins.*, 75:594–612, 2004.
[23] H. Flyvbjerg. *unpublished*, 2003.
[24] A. S. Grove. *Physics and Technology of Semiconductor Devices*. John Wiley and Sons, New York, 1967.
[25] W. H. Press, B. P. Flannery, S. A. Teukolsky, and W. T. Vetterling. *Numerical Recipes*. Cambridge University Press, 1986.
[26] Henrik Flyvbjerg and Henrik Gordon Petersen. *J. Chem. Phys.*, 91:461–466, 1989.
[27] N. C. Barford. John Wiley & Sons, 2nd edition, 1986.
[28] Photodiode Characteristics, UDT Sensors Inc., 2002. www.udt.com.
[29] Kirstine Berg-Sørensen, Erwin J. G. Peterman, Lene Oddershede, Meindert van Dijk, Ernst-Ludwig Florin, Christoph F. Schmidt, and Henrik Flyvbjerg. *Proceedings of SPIE*, 5514:419–427, 2004.